\documentclass[12pt,letterpaper]{article}
\usepackage{amsmath}
\usepackage{amssymb}
\usepackage{cite}
\usepackage{dcolumn} %% tables cols aligned at decimal point
\usepackage{setspace}
\usepackage{array}
\usepackage{psfrag}
\usepackage{graphicx}
\usepackage{endfloat}
\def\brp{{\mathbf{r}^{\prime}}}
\def\bxp{{\mathbf{x}^{\prime}}}
\def\tp{{{t}^{\prime}}}

\def\br{{\mathbf{r}}}
\def\bx{{\mathbf{x}}}
\def\bR{{\mathbf{R}}}

\def\d{{\mathrm{d}}}
\def\rhor{{\rho({\bf r})}}

\def\rhoi{{\rho_I}}
\def\rhoj{{\rho_J}}
\def\rhoir{{\rho_I({\bf r})}}

\def\rhojrp{{\rho_J({\bf r}^{\prime})}}
\def\sumi{{\sum_I^{N_S}}}

\def\im{{\operatorname{Im}}}

\def\vdw{{van der Waals}}

\newcommand{\eqn}[1]{\mbox{Eq.\hspace{1pt}(\ref{#1})}}
\newcommand{\eqs}[2]{\mbox{Eq.\hspace{1pt}(\ref{#1}--\ref{#2})}}
\newcommand{\pot}[1]{v_{\rm #1}}
\begin{document}

\begin{center}
\vspace*{1cm}
{\LARGE\bf FDE-vdW: A van der Waals Inclusive Subsystem Density-Functional Theory}\\[3ex]

{\large Ruslan Kevorkyants,$^1$  Henk Eshuis,$^2$ and Michele Pavanello$^1$}\\
$^1$Department of Chemistry, Rutgers University, Newark, NJ 07102, USA\\
$^2$Department of Chemistry and Biochemistry, Montclair State University, Montclair, NJ 07043, USA\\[2ex]
\end{center}

\vfill

\begin{tabbing}
Date:   \quad\= \today \\
Status: \> Submitted to J.\ Chem.\ Phys.\ \\
\end{tabbing}
\newpage
\begin{abstract}
We present a formally exact van der Waals inclusive electronic structure theory, called FDE-vdW, 
based on the Frozen
Density Embedding formulation of subsystem Density-Functional Theory. 
In subsystem DFT, the energy functional is composed of subsystem additive and
non-additive terms. We show that an appropriate definition of the long--range correlation
energy is given by the value of the non-additive correlation functional. 
This functional is evaluated using the Fluctuation--Dissipation Theorem aided by a formally exact decomposition
of the response functions into subsystem contributions.
%, van der Waals corrected binding
%energies of weakly bound molecular systems can be computed. Theory side, we analyze
FDE-vdW is derived in detail and several approximate schemes are proposed, which lead to practical
implementations of the method. We show that FDE-vdW is Casimir--Polder consistent,
i.e.\ it reduces to the generalized Casimir--Polder formula for asymptotic
inter-subsystems separations. Pilot calculations of binding energies of 13 weakly bound
complexes singled out from the S22 set
show a dramatic improvement  upon semilocal
subsystem DFT, provided that an appropriate exchange functional is employed.
The convergence of FDE-vdW with basis set size is discussed, as well as its dependence on
the choice of associated density functional approximant.
\end{abstract}
\newpage
\section{Introduction}
In subsystem Density-Functional Theory \cite{jaco2013} (subsystem DFT), the total electronic density of a molecular system is partitioned into subsystem contributions:
\begin{equation}
\label{int1}
\rho(\br) = \sumi \rhoi(\br),
\end{equation}
where $N_S$ is the total number of subsystems. This is an appealing strategy, as it invokes a kind of ``divide and conquer'' principle which makes the electronic problem more manageable by partitioning it into several coupled sub-problems. This strategy is general, and can be formulated in such a way as to involve no approximations \cite{cohe2007a,grit2013}. In practice, this divide and conquer approach is more easily implemented if the electron system is composed of weakly interacting subsystems \cite{weso2006,weso1993}. In this limiting case, the total electron density in \eqn{int1} is already well approximated by using the electron densities of isolated subsystems (i.e.\ the molecular fragments composing the full system). 

Subsystem DFT is designed (see Section 2) to reproduce Kohn--Sham DFT \cite{hohe1964}  (KS-DFT), and thus is formally exact \cite{weso2006}. Because of this, subsystem DFT inherits all the strengths and weaknesses of KS-DFT, e.g.\  the fact that the exact exchange--correlation (XC) functional, $E_{\rm xc}[\rho]$, is unknown. As a consequence, practical implementations of subsystem and KS-DFT involve employing
XC functional approximants. The most common class of approximants rely on local and semilocal forms of the functional, i.e.\ it depends on the electron
density 
and possibly also on its  gradient at a
point $\br$ in space. 
Due to their character, these semilocal approximations 
are unable to account for interactions that are non-local in nature, such as dispersion
interactions \cite{mcla1963,long1965} and generally all long--ranged interactions
originating from the correlated part of the energy functional. 

Several solutions have been proposed to address this deficiency of local and semilocal
XC functionals~\cite{klim2012}. One avenue is the design of fully non-local functionals,
as proposed by Dion {\em et al.}~\cite{lang2005} and by Vydrov and Van
Voorhis~\cite{vydr2010a}. A very popular and efficient 
alternative has been an additive pairwise correction that uses the so-called $C_6$ coefficients \cite{sun2008,grim2004,elst2001,jure2007,sato2005,vonl2004,sato2010,beck2007,kris2012,stei2011}. This type of correction is 
inspired by the Casimir-Polder formula,
\begin{equation} 
\label{CP}
E_{\rm disp} = \frac{C_6^{\rm AB}}{R_{\rm AB}^6},
\end{equation}
where $R_{\rm AB}$ is the distance between fragment $A$ and $B$ in a molecular system. 
Appropriate decompositions of the molecular system  are used to yield 
pairwise corrections \cite{grim2010,tkat2012,rob2013}. 

A better description of 
dispersion interactions is given by the generalized Casimir-Polder (GCP) formula
\cite{mcla1963,long1965}
\begin{equation}
\label{fdt8}
E_{\rm disp}=-\frac{1}{2\pi}\int_0^{+\infty}\d\omega \int \frac{\chi_1(\brp_1,\br_1,i\omega)\chi_2(\br_2,\brp_2,i\omega)}{|\br_1-\br_2|~|\brp_1-\brp_2|}\d\br_1\d\br_2\d\brp_1\d\brp_2,
\end{equation}
where $\chi_1$ and $\chi_2$ are the electronic frequency dependent linear-response
functions of the two isolated fragments. Here, the fragment's electron
densities are allowed to overlap and the restriction to large interfragment distance is
also lifted. The GCP formula is derived to account for
dispersion interactions to second order in perturbation theory, where the perturbation is
the Coulomb interaction between the fragments.

A formally exact framework to obtain the full correlation energy (as
opposed to only the dispersion part) is the adiabatic--connection 
fluctuation-dissipation theorem applied to DFT
(ACFDT-DFT) \cite{furc2005,kohn1998,dobs2012} which relates the total correlation energy to
the frequency dependent linear-response functions (response functions, hereafter). 
%in a way
%that is qualitatively similar to GCP. 
Since  response functions can be efficiently calculated using 
semilocal DFT \cite{misq2002,hess2002,casi1995,furc2001,furc2005}, ACFDT-DFT is
a very appealing way to introduce dispersion interactions, and generally all interactions
arising from the correlated part of the energy. 

The simplest and most used method based
on the ACFDT
is the random phase approximation (RPA) \cite{furc2008,eshu2012,Ren2012, nguy2009, nguy2010}
and its extensions~\cite{Bates2013,Paier2012}. 
RPA is parameter free, and was shown to be an order of magnitude more accurate than semi-local DFT for mid- and
long-range correlation~\cite{Eshuis2011} and of comparable accuracy as dispersion-corrected
functionals, such as B3LYP-D3, or highly parameterized functionals, such as M06-2X \cite{zhao2006b}. 
For short-range correlation, RPA does not improve over semi-local DFT, thus
range-separated schemes were proposed which combine a short-range DFT description of the
system with a long-range RPA description. One advantage of these methods is a much faster
convergence with basis set size at the cost of having to include a parameter in the formalism \cite{brun2012,toul2009,toul2004}. 

The main idea behind this work is to devise a theory that exploits the fragment
picture provided by subsystem DFT and borrows ideas from ACFDT-DFT to obtain the (theoretically exact) correlation energy of
interaction between fragments. This has, so far, not been  attempted for overlapping electronic
fragment densities (for non-overlapping densities see Ref.\ \citenum{dobs2012}). 
As we exploit the Frozen Density Embedding (FDE) formulation of
subsystem DFT, we call the resulting theory FDE-vdW. 

FDE-vdW has several appealing properties: (1) full correlation energies
are recovered, (2) the theory is formally exact and 
independent of the definition of the fragments (or subsystems), (3)
the long-range correlation
energy is rigorously incorporated in the subsystem DFT energy functional without resorting to an ad-hoc function that switches between the short and the long range.

Two recent publications provide a major motivation for this work. First, Visscher and coworkers \cite{beyh2013}, demonstrated that the FDE coupled with semiempirical dispersion corrections yields accurate interaction energies. Second, Cha\l{}asi\'nski and coworkers \cite{luka2010}, used a projector-based partitioning of HF and KS-DFT wavefunctions into fragment wavefunctions to define an inter-fragment dispersion correction calculated with the GCP formula which yielded excellent interaction energies. In both of the mentioned publications as well as in this work, the common double-counting of van der Waals interactions occurring when correcting the KS-DFT interaction energies is completely bypassed and the resulting theories are more balanced.

The paper is organized as follows. Section 2 recalls some basic ideas of
subsystem DFT and its time-dependent extension. Section 3 develops the FDE-vdW
theory by applying ACFDT-DFT to subsystem DFT. Section 4  gives an outline of
possible approximations for  practical calculations. Section 5 presents
details of the FDE-vdW implementation. Section 6 displays pilot calculations carried out
with one of the approximations proposed in Section 4. Finally, Section 7 delineates conclusions and
future directions for the presented formalism.

\section{Subsystem DFT and TD-DFT}

The density partitioning in \eqn{int1} is in principle arbitrary. One can
use chemical intuition to identify subsystems (using for example a
real-space partitioning), or any other criterion. Unlike the
density, the energy functional is not additive, and an exact form
of it involves non-additive corrections that arise at the
kinetic-energy and XC energy level as well as a Hartree term
describing the electrostatic interaction between subsystems,
\begin{equation}
\label{int2}
E\left[\sumi \rhoi\right] = \sumi E[\rhoi] + J^{\rm nadd}[\{\rhoi\}]+ T^{\rm nadd}_s [\{\rhoi\}]+E^{\rm nadd}_{xc} [\{\rhoi\}],
\end{equation}
where $E[\rhoi]=T_s[\rhoi]+J[\rhoi]+E_{xc}[\rhoi]+V_{\rm
ext}[\rhoi]$, and in $J$ we have grouped the nuclear attraction
term. A computationally attractive way to find subsystem densities
\cite{weso1993} is the Frozen Density Embedding theory (FDE). In FDE the energy functional of all the subsystem densities is minimized  in
\eqn{int2} w.r.t.\ only one subsystem density while keeping all
the others frozen. This procedure can be repeated until self-consistency \cite{weso2003}. Freezing all densities but one during
the minimization, is equivalent to using partial functional
derivatives in the working KS equations and with that, mixed
derivatives such as 
$\frac{\delta E[\rhoj]}{\delta\rhoi}$ are always set to zero in
FDE.

The aim of FDE is to represent the subsystems as a set of $N_S$
coupled Kohn--Sham systems. Hence, the subsystem densities must be
non-negative, must integrate to a preset number of electrons,
i.e.\ $\int\rhoir\d\br=N_I$, and must be non-interacting $v$-representable. In
practical FDE calculations, the subsystem densities are
constructed from subsystem molecular orbitals which are expanded
in terms of localized atomic orbitals, often centered on atoms
belonging to only one molecular fragment in a system (monomer
basis set). Thus, the subsystems are identified as non-covalently
bound molecules. 
Recently, however, embedding schemes based on optimized effective potential \cite{good2010,good2011a,huang2011}, projection operators \cite{luka2010,khai2012,barn2013,manb2012,good2012}, and fitted functionals \cite{gill2013,hu2012}, are paving the way for the correct description of covalent bonds (high density overlap) by FDE.

Self-consistent solution of a set of coupled KS-like equations (also called KS equations with constrained electron density \cite{weso2006}) yields the set of subsystem KS orbitals
\begin{equation}
\label{ks1}
\left[ -\frac{1}{2} \nabla^2 + \pot{eff}^I(\br) \right] \phi^I_k(\br) = \varepsilon^I_k  \phi^I_k(\br),\mathrm{~with~}I=1,\ldots,N_S
\end{equation}
with the effective subsystem potential given by
\begin{equation}
\label{sks}
\pot{eff}^I (\br)=\underbrace{\pot{eN}^I (\br) + \pot{Coul}^I(\br) +\pot{xc}^I (\br)}_{\mathrm{same~as~regular~KS-DFT}} + \pot{emb}^I (\br).
\end{equation}
In the FDE formulation of subsystem DFT \cite{weso1993,weso2006}, the potential above, $\pot{emb}$, is called embedding potential and is given by
\begin{align}
\label{emb}
\nonumber
\pot{emb}^I (\br)&=\sum_{J\neq I}^{N_S} \left[ \int \frac{\rhojrp}{|\br-\brp|} \d\brp - \sum_{\alpha\in J} \frac{Z_\alpha}{|\br-\bR_{\alpha}|} \right] + \\
&+\frac{\delta T_s[\rho]}{\delta\rhor}-\frac{\delta T_s[\rhoi]}{\delta\rhoir}+\frac{\delta E_{\rm xc}[\rho]}{\delta\rhor}-\frac{\delta E_{\rm xc}[\rhoi]}{\delta\rhoir}.
\end{align}

In the case of closed-shell subsystems, the density of the supersystem is thus found using \eqn{int1} and \eqn{ks1} as 
\begin{equation}
\label{rho}
\rhor=2\sumi\sum_i^{{\rm occ}_I} \left| \phi^I_i(\br)\right|^2.
\end{equation}

The time-dependent extension of subsystem DFT \cite{casi2004,neug2010a,neug2009,neug2007} is based on the Runge-Gross theorem \cite{rung1984} which was recently proved also for subsystem DFT by Wasserman {\em et al.} \cite{mosq2013}. Recently, linear-response subsystem TD-DFT was formulated in terms of subsystem response functions by one of us \cite{pava2013b}. The relevant Dyson-type equations for the subsystem response functions read as follows (contracting the integrals in a short-hand notation)
\begin{equation}
\label{dyson}
\chi_I^c= \chi_I^u + \sum_{J\neq I}^{N_S} \chi_I^u K_{IJ} \chi_J^c,
\end{equation}
where the correlated {\bf coupled} subsystem response function, $\chi_I^c$, is given by
\begin{equation}
\label{resp}
\chi_I^c = \frac{\delta \rho_I}{\delta \pot{ext}},
\end{equation}
and the correlated {\bf uncoupled} subsystem response function, $\chi_I^u$, is found by solving the Dyson equation
\begin{equation}
\label{dysonu}
\chi_I^u= \chi_I^0 + \chi_I^0 K_{II} \chi_I^u.
\end{equation}
In the above equation, $\chi_I^0$ is the response of the KS subsystem. The kernel matrix, $K_{IJ}$, in \eqn{dyson} and \eqn{dysonu} is defined as
\begin{equation}
\label{kernel}
K_{IJ}(\br,\brp,\omega)=\frac{1}{| \br - \brp |} + f_{\rm xc}(\br,\brp,\omega)+ f_{\rm T}(\br,\brp,\omega) - f_{\rm T}^I(\br,\brp,\omega) \delta_{IJ},
\end{equation}
where the kinetic energy kernels, expressed in the time domain, are defined as
\begin{align}
\label{kinkt}
f_{\rm T}(\br,\brp,t-\tp)&=\frac{\delta^2 T_{\rm s}[\rho]}{\delta \rho(\br,t) \delta\rho(\brp,\tp)},\\
\label{kinks}
f_{\rm T}^I(\br,\brp,t-\tp)&=\frac{\delta^2 T_{\rm s}[\rhoi]}{\delta \rhoi(\br,t) \delta\rhoi(\brp,\tp)}.
\end{align}
\section{Fluctuation--Dissipation Theorem Subsystem DFT}
The correlation energy within the Adiabatic-Connection
Fluctuation--Dissipation Theorem DFT (ACFDT-DFT) can be written exactly as
\begin{equation}
\label{fdt1}
E_{c}=\int\int \d \bx_1 \d \bx_2
\frac{P_c(\bx_1,\bx_2)}{|\br_1 - \br_2|}
\end{equation}
where $\bx_i=\{\br_i, \sigma_i \}$ is the $i$-th spin-spatial coordinate, and $P_c(\bx_1,\bx_2)$ is the correlated part of the electronic pair density which is defined as \cite{furc2005,dobs2012}
\begin{equation}
\label{fdt2}
P_c(\bx_1,\bx_2)=-\frac{1}{2\pi} \im\left[ \int_0^1 \d \alpha \int_0^{+\infty} \d \omega 
\left( \chi^\alpha(\bx_1,\bx_2,\omega)-\chi^0(\bx_1,\bx_2,\omega) \right) \right].
\end{equation}
The functions $\chi^0$ and $\chi^\alpha$ are the electron's
Kohn-Sham and interacting response functions, respectively. The
interacting response is evaluated at coupling strength $\alpha$
and the adiabatic connection formula is used
\cite{dobs2012,gunn1976}. 
The central problem is finding analytical and/or
computationally appealing ways to 
evaluate \eqs{fdt1}{fdt2}. 

As the response functions explicitly depend on both occupied and
virtual orbitals, for RPA and TD-DFT alike, a formal $N^6$ scaling
with system size arises in their computation. For realistic systems this leads to a large
computational expense and it is usually handled employing numerical techniques based on fitting of the response functions or the entire correlation energy expression \cite{eshu2010,nguy2009,pode2012}.
An important aspect of this work is that the computation of 
\eqs{fdt1}{fdt2} for the supersystem is completely avoided. 
We elaborate on this crucial aspect in the next section.

\subsection{Short range, subsystem additive correlation energy}
We define the short range, intra-subsystem correlation energy as the correlation energy derived with \eqn{fdt1} with the pair density including only the response functions calculated in the uncoupled formalism of \eqn{dysonu}. This definition is inspired by the idea that when the subsystems are separated by an infinitely large distance, we have $\chi_I^c=\chi_I^u=\chi_I$ (where $\chi_I$ is the response function of the isolated subsystem). In this asymptotic case, the short range correlation energy {\it depends only on the degrees of freedom of a single subsystem}, and the correlated part of the interaction between subsystems vanishes. 

According to our definition, the additive (or short range) part of the correlated pair density, $P_c^{\rm add}$, to be used in \eqn{fdt2} depends on the difference of response functions
\begin{equation}
 \label{fdt4}
\Delta\chi_\alpha^{\rm add} =  \left(\sumi \chi^{u,\alpha}_I\right) -\chi^0 \simeq \sumi \left(  \chi^{u,\alpha}_I -\chi_I^0 \right),
\end{equation}
where we have used the approximation $\sumi \chi_I^0\simeq\chi^0$, which was discussed previously \cite{pava2013b}.
To obtain the above uncoupled subsystem response function one
needs to use \eqn{dysonu} and evaluate the interaction kernel
$K_{IJ}^\alpha$
at coupling strength $\alpha$,  
\begin{equation}
\label{kernela}
K_{IJ}^\alpha(\br,\brp,\omega)=\frac{\alpha}{| \br - \brp |} + f^\alpha_{\rm xc}(\br,\brp,\omega)+ f^\alpha_{\rm T}(\br,\brp,\omega) - f^{I,\alpha}_{\rm T}(\br,\brp,\omega) \delta_{IJ}.
\end{equation}
Even though the additive correlation energy can theoretically be calculated with this formulation, it is not computationally advantageous \cite{eshu2012,eshu2010} and poses problems related to a singular $P_c^{\rm add}(\bx_1,\bx_2)$ as $\bx_1\to\bx_2$ due to the (semi) locality of commonly adopted XC functionals \cite{furc2005,kimb1976}. Hereafter, we will make the assumption that the additive (or short-range) correlation is well captured by semilocal XC density functional approximants. Thus, the remainder of this work focuses on the inter-subsystem correlation energy, by finding an appropriate expression for the long range (or non-subsystem-additive) part of the correlated pair density, $P_c^{\rm nadd}(\bx_1,\bx_2)$.
\subsection{Long range non-additive correlation energy}
The subsystem DFT density functional in \eqn{int2} is comprised of subsystem-additive and non-additive terms. Specifically, the non-additive XC functional can be further split into exchange and correlation contributions, $E_{\rm xc}^{\rm nadd}=E_{\rm x}^{\rm nadd}+E_{\rm c}^{\rm nadd}$. The long range contribution to the correlation energy is defined as the energy term left subtracting the short-range contribution from the total correlation energy,
\begin{equation}
\label{eq:splitcorr}
E_{\rm c}=\underbrace{\sumi E_{\rm c}\left[ \rhoi \right]}_{\mathbf{\mathrm{ Short~Range }}} + \underbrace{E_{\rm c}^{\rm nadd}\left[ \rho_I, \rho_{II},\ldots,\rho_{N_S} \right]}_{\mathbf{\mathrm{ Long~Range }}}.
\end{equation}
The difference of response functions that yields the non-additive part of the correlated pair density is
\begin{align} 
\nonumber
\Delta\chi_\alpha^{\rm nadd} &=\sumi \left(  \chi^{c,\alpha}_I  -
\chi^{u,\alpha}_I \right)\\
\label{django}
&= \sumi \Delta\chi_{\alpha,I}^{\rm nadd},
\end{align}
which by inspection of \eqn{dyson} is non-subsystem-additive. Thus, the long range correlation energy derived from \eqn{django} will naturally replace the non-additive correlation energy functional in \eqn{int2}. It is easily shown that 
\begin{equation}
\label{eq:totalsum}
\chi^\alpha-\chi^0 = \Delta\chi_\alpha^{\rm add} + \Delta\chi_\alpha^{\rm nadd}.
\end{equation}
In order to obtain the non-additive correlated pair density, integrals in both the coupling strength ($\alpha$) and frequency ($\omega$) of the difference of subsystem response functions defined above need to be evaluated
\begin{equation}
\label{sublr1}
P_c^{\rm nadd}(\bx_1,\bx_2)=-\frac{1}{2\pi}\im\left[ \int_0^1 \d \alpha \int_0^{+\infty} \d \omega 
\sumi \Delta\chi_{\alpha,I}^{\rm nadd}(\omega) \right].
\end{equation}
Thus, the long range component of the correlation energy can be
calculated substituting \eqn{sublr1} into \eqn{fdt1}.
\eqn{sublr1} is the key result of this work. 
Employing \eqn{dyson} in \eqs{django}{sublr1} results in
\begin{equation}
\label{sublr3}
E_{c}^{\rm nadd}=-\frac{1}{2\pi}\im\left[\int\int \d \bx_1 \d \bx_2
\frac{ \int_0^1 \d \alpha \int_0^{+\infty} \d \omega 
\sumi \sum_{J\neq I}^{N_S} \chi_I^{u,\alpha}(\omega) K_{IJ}^\alpha(\omega)\chi_J^{c,\alpha}(\omega) }{|\br_1 - \br_2|}\right].
\end{equation}
\section{Approximate treatments of the non-additive correlation energy}
The non-additive correlation energy determined with \eqn{sublr3}
is still computationally prohibitive for realistic systems, 
as the coupled response functions
must be obtained solving the Dyson-type equation in \eqn{dyson}
which in theory requires $N^6$ operations, where $N$ is the size
of the entire supersystem \cite{pava2013b,koni2012b}. It is, therefore, important to develop
simplified expressions to compute the non-additive correlation
energy. The first approximation is obtained 
by considering a perturbative solution to the Dyson-type equation \eqn{dyson} as
\begin{equation}
\label{sublr4}
\chi_I^{c,\alpha} - \chi_I^{u,\alpha} \simeq \sum_{J\neq I}^{N_S} \chi_I^{u,\alpha} K_{IJ}^{\alpha} \chi_J^{u,\alpha}+\sum_{{ J\neq I,~ K\neq J }}^{N_S} \chi_I^{u,\alpha} K_{IJ}^{\alpha} \chi_J^{u,\alpha} K_{JK}^{\alpha} \chi_K^{u,\alpha}+ \cdots
\end{equation}
and retaining only the first term of the expansion. \eqn{sublr3}
then simplifies to
\begin{equation}
\label{sublr3b}
E_{\rm c}^{\rm nadd}=-\frac{1}{2\pi}\im\left[ \int\int \d \bx_1 \d \bx_2
\frac{\int_0^1 \d \alpha \int_0^{+\infty} \d \omega 
\sumi \sum_{J\neq I}^{N_S} \chi_I^{u,\alpha}(\omega) K_{IJ}^\alpha(\omega) \chi_J^{u,\alpha}(\omega) }{|\br_1 - \br_2|}\right].
\end{equation}
The non-additive correlation energy is now
expressed in terms of the functions $\chi_I^u$ (i.e.\ the uncoupled subsystem response functions).
In practice \eqn{sublr3b} has to be further approximated since $K_{IJ}$
is not known exactly. Motivated by the success of RPA for non-covalent
interactions, we approximate $K_{IJ}$ by neglecting the
frequency-dependent exchange-correlation kernel as well as the
kinetic energy contributions to the kernel ($f_{\rm xc}$, $f_{\rm T}$, and $f_{\rm T}^I$)
\begin{equation}
\label{kernel_rpa}
K_{IJ}^{\alpha } \approx 
K_{IJ}^{\alpha, \text{RPA}} = 
 \frac{\alpha}{|\br-\brp|}.
\end{equation}
Since the RPA kernel is frequency independent, it is computationally
much more efficient to solve \eqn{sublr3b}. We dub the approximate
method GCP$_{\alpha}^u$. One can reduce the computational cost further by
applying subsequent approximations. We obtain GCP$_{1}^u$ by
evaluating the coupling strength integration for subsystems evaluated at
$\alpha =1$.
Finally, a further approximation can be carried out (although we never do in this work) by which the subsystems can be treated as isolated (GCP). 
We summarize the various approximations and their proposed acronyms
in Table \ref{tappr}.
\begin{table}
\begin{center}
\begin{tabular}{m{3.5cm}cc}
{\sc Description} & {\sc Integrand of \eqn{sublr3b}} & {\sc Acronym}\\
\hline\\[-8pt]
Exact & $\chi_I^{u,\alpha}K_{IJ}^\alpha\chi_J^{c,\alpha}$ & \\[10pt]
\begin{minipage}{3.5cm}{RPA \\ \& Perturbative}\end{minipage} &
$\chi_I^{u,\alpha}\frac{\alpha}{|\br-\brp|}\chi_J^{u,\alpha}$ &
GCP$_\alpha^u$\\[25pt]
\begin{minipage}{3.5cm}{$\chi_{I/J}^{u,\alpha=1}$\\ \& RPA \\ \&
Perturbative}\end{minipage} &
$\chi_I^{u}\frac{\alpha}{|\br-\brp|}\chi_J^{u}$ & GCP$^{\rm
u}_{1}$\\[30pt]
\begin{minipage}{3.5cm}{Isolated-fragment response functions \\ \& RPA
\\ \& Perturbative}\end{minipage} &
$\chi_I\frac{\alpha}{|\br-\brp|}\chi_J$ & GCP\\[10pt]
\end{tabular}
\end{center}
\caption{\label{tappr} Proposed approximations to \eqn{sublr3b} and their respective acronym. ``RPA'' stands for Random Phase Approximation and it is applied anytime the kernel $K_{IJ}$ is depleted of the XC and kinetic terms. ``GCP'' stands for Generalized Casimir-Polder formula.}
\end{table} 

For only two interacting subsystems ($N_S=2$),
application of the GCP$^u_{1}$ approximation and subsequent 
integration over $\alpha$ (which gives a factor $\frac{1}{2}$) yields
\begin{equation}
\label{subecorr}
E_{\rm c}^{\rm nadd}=-\frac{1}{2\pi}\im\left[\int\int \d \bx_1 \d \bx_2
\frac{\int \d\bx\d\bxp  \int_0^{+\infty} \d \omega \chi_1^{u}(\bx_1,\bx,\omega) \frac{1}{|\br-\brp|}\chi_2^{u}(\bxp,\bx_2,\omega)}{|\br_1 - \br_2|}  \right].
\end{equation}
The use of
 a coupling-strength independent response function is
analyzed in detail and justified in the appendix section.

An important test of the meaningfulness of the additive and
non-additive correlation energy definitions given in this work can
now be carried out by analyzing the behavior of the non-additive
correlation in the limit of large subsystem separations (or
equivalently negligible intersubsystem density overlap). In this
limit, the correct espression of the \vdw\ interaction is the
 generalized Casimir-Polder or Longuet-Higgins formula
(first derived by McLachlan \cite{mcla1963} for molecules) given
in \eqn{fdt8} of the introduction section. The GCP formula is
analogous to \eqn{subecorr} and  is recovered by adopting the
``GCP'' approximation listed in Table \ref{tappr}. As mentioned
before, for large inter-subsystem separations,
$\chi_I^{c}\simeq\chi_I^{u}\simeq\chi_I$. Thus, in the asymptotic
limit, this theory reduces to the GCP formula -- which is an
important property, also termed {\it Casimir-Polder consistency}
\cite{neepa_book}. The GCP formula has a long history, both in
wavefunction-based and DFT methods for correcting the mean-field approximation
\cite{misq2003,misq2002,jezi1994,hess2003,hess2002b}. 
\section{Practical Calculations}
\subsection{Integration of the fluctuation-dissipation formula}

For closed--shell systems, the spectral representation of the response functions, assuming real solutions of the TD-DFT eigenvectors, is given by \cite{furc2001}
\begin{eqnarray}
\label{tddft_pc}
\chi_I^u(\br,\brp,\omega) = \sum_{(n)_I} \frac{4\omega_n^u}{(\omega_n^u)^2 -\omega^2}\sum_{(ia)_I,(jb)_I} (X^u_n+Y^u_n)_{ia}(X^u_n+Y^u_n)_{jb} \\
\label{sublr5}
\phi_{i}(\br) \phi_{a}(\br) \phi_{j}(\brp)\phi_{b}(\brp),
\end{eqnarray}
where $\phi_{i}(\br)$ are the KS orbitals of subsystem $I$. Subscripts
$i,j,k,l$ denote occupied orbitals and $a,b,c,d$ virtual orbitals. 
$(X^u_n+Y^u_n)_{ia}$ is the projection of the sum of  the excitation
($X$) and de-excitation ($Y$) TD-DFT eigenvector for the $n$-th
excited state of subsystem $I$. The associated eigenvalue is given by
$\omega_n^u$.  The Hartree--XC kernel $K_{II}$
used to determine $\chi_I^u$ is given by \eqn{kernel}
with $I=J$ \cite{casi2004,neug2005b}. 
Using \eqn{tddft_pc} and employing the GCP$_1^u$
approximation (see Table \ref{tappr}) we can express
the difference of the coupled and uncoupled response functions
\begin{align}
\label{sublr6}
\nonumber
&\chi_I^{c,\alpha}(\br_1,\br_2,\omega) - \chi_I^{u,\alpha}(\br_1,\br_2,\omega) \simeq \\
\nonumber
&\sum_{J\neq I}^{N_S}\sum_{(n)_I}^{o_Iv_I}\sum_{(m)_J}^{o_Jv_J}\sum_{(ia)_I}^{o_Iv_I}\sum_{(jb)_I}^{o_Iv_I}\sum_{(kc)_{J}}^{o_Jv_J}\sum_{(ld)_{J}}^{o_Jv_J}\frac{16\omega_n^u\omega_m^u}{\big((\omega_n^u)^2 -\omega^2\big)\big((\omega_m^u)^2 -\omega^2\big)}\\
\nonumber
&(X^u_n+Y^u_n)_{ia}(X^u_n+Y^u_n)_{jb}(X^u_m+Y^u_m)_{kc}(X^u_m+Y^u_m)_{ld}\\
&\phi_{j}(\br_1)\phi_{b}(\br_1)\phi_{k}(\br_2) \phi_{c}(\br_2)\int \d\br\d\brp \frac{\alpha}{|\br-\brp|}\phi_{i}(\br) \phi_{a}(\br)  \phi_{l}(\brp)\phi_{d}(\brp).
\end{align}
%The coupling strength integration yields a factor $\frac{1}{2}$.
Integrating over coupling strength and 
taking into account double
counting arising from the summation over the subsystems,
we obtain for \eqn{sublr3}
\begin{eqnarray}
\label{sublr7}
\nonumber
E_{\rm c}^{\rm nadd}=-\frac{8}{\pi}\sumi  \sum_{J\geq I}^{N_S} \sum_{(n)_I}^{o_Iv_I}\sum_{(m)_J}^{o_J+v_J}  \int_0^{+\infty} \d \omega\frac{\omega_n^u\omega_m^u}{\big((\omega_n^u)^2 +\omega^2\big)\big((\omega_m^u)^2 +\omega^2\big)}  \\
\sum_{(ia)_I}^{o_Iv_I}\sum_{(jb)_I}^{o_Iv_I}\sum_{(kc)_{J}}^{o_Jv_J}\sum_{(ld)_{J}}^{o_Jv_J}(X^u_n+Y^u_n)_{ia}(X^u_n+Y^u_n)_{jb} (X^u_m+Y^u_m)_{kc}(X^u_m+Y^u_m)_{ld}\\
\nonumber
\langle ia | ld \rangle \langle jb | kc \rangle.
\end{eqnarray}
Finally, the frequency integration can be carried out analytically \cite{furc2008},
\begin{eqnarray}
\label{sublr8}
\nonumber
E_{\rm c}^{\rm nadd}=\sumi  \sum_{J\geq I}^{N_S} \sum_{(n)_I}^{o_Iv_I}\sum_{(m)_J}^{o_Jv_J} \frac{4}{(\omega_n^u +\omega_n^u)} \sum_{(ia)_I}^{o_Iv_I}\sum_{(jb)_I}^{o_Iv_I}\sum_{(kc)_{J}}^{o_Jv_J}\sum_{(ld)_{J}}^{o_Jv_J}  \\
(X^u_n+Y^u_n)_{ia}(X^u_n+Y^u_n)_{jb} (X^u_m+Y^u_m)_{kc}(X^u_m+Y^u_m)_{ld}\\
\nonumber
\langle ia | ld \rangle \langle jb | kc \rangle.
\end{eqnarray}

\subsection{Implementation of the GCP$_\alpha^u$ method}
\eqn{sublr8} was implemented  in the {\sc Adf} computer software \cite{teve2001a} as a modification of the existing SUBEXCI code \cite{koni2011,koni2012b,neug2010a,neug2007}. The SUBEXCI code uses the so-called $Z$-vector (Hermitian) formulation of the TD-DFT equations. \eqn{sublr8} is transformed to
\begin{equation}
\label{sublr6b}
E_{\rm c}^{\rm nadd}=-4\sumi  \sum_{J\geq I}^{N_S}\sum_{(n)_I(m)_J} \frac{\big|\Omega_{nm}^{IJ}\big|^2}{\omega_n\omega_m(\omega_n+\omega_m)}.
\end{equation}
In the above, 
\begin{equation}
\label{sublrADF}
\Omega_{nm}^{IJ}=\left(S^{-\frac{1}{2}} Z_n\right)^T K_{IJ} \left(S^{-\frac{1}{2}}Z_m\right),
\end{equation}
where $S_{ia}=\varepsilon_{a}-\varepsilon_i$, the $Z$ vectors are the solutions of the Hermitian TD-DFT equations and, following the GCP$^u_\alpha$ approximation, $K_{IJ}$ in \eqn{sublrADF} is modified to only include the Coulomb kernel.

As {\sc Adf} is a Slater-Type Orbital code, two-electron integrals are not available. For this reason, the $\Omega_{nm}$ matrix elements are calculated in density fitting. For each matrix element, the inducing density $Z_m$ is fitted to yield a fitted induced potential, $K_{IJ}Z_m$. The induced potential is then integrated over a monomer-based integration grid with $Z_n$. This is not a very efficient code for our purposes. In the recent works by Szalewicz {\it et al.} \cite{pode2012} regarding the implementation of the GCP formula, density fitting is applied to the entire response function before carrying out the frequency integration. Eventually, our {\sc Adf} code will include a Szalewicz-type density fitting. However, as this work constitutes a proof-of-principle, we have left additional coding for a future work.
\subsection{Binding energy combining the non-local correlation with the FDE energy: The FDE-vdW method}
In FDE, the total binding energy of two subsystems can be calculated by subtracting from the total FDE energy the energy of the isolated subsystems \cite{gotz2009}. In the case of only two subsystems, we have
\begin{align}
\nonumber
\label{FDEbind}
E_{\rm bind}^{\rm FDE} &= \underbrace{\left( E[\rho_I] + E[\rho_{II}] \right) - \left( E[\rho_I^{\rm iso}] + E[\rho_{II}^{\rm iso}] \right)}_{\mathrm{{\bf Orbital~Relaxation}}}
+\underbrace{T^{\rm nadd}[\rho_I,\rho_{II}]+E_{\rm xc}^{\rm nadd}[\rho_I,\rho_{II}]}_{\mathrm{{\bf Non-additive~T~and~XC}}} \\
                &+\underbrace{\int \frac{\rho_I(\br)\rho_{II}(\br^\prime)}{|\br-\br^{\prime}|}\d\br\d\br^\prime + \int v^I_{\rm eN}(\br)\rho_{II}(\br)\d\br+ \int v^{II}_{\rm eN}(\br)\rho_{I}(\br)\d\br}_{\mathrm{{\bf Coulomb/Polarization}}},
\end{align}
where we have introduced $\rho_{I/II}^{\rm iso}$ indicating the electron density of the isolated subsystem $I/II$. Pernal {\it et al.} \cite{pern2009b} have devised a strategy to first remove the dispersion interactions from the density functional, and then to augment the obtained binding energy by the non-local dispersion interaction obtained from the GCP formula. Rajchel {\em et al.} \cite{luka2010}, instead, are able to compute dispersionless interactions by partitioning the HF or KS wavefunctions into fragment wavefunctions. Here we largely follow these ideas. However, instead of reparametrizing the density functional to remove the dispersion interactions from it, after the FDE procedure has completed and subsystem densities are recovered, we calculate the FDE energy omitting the correlation part of the non-additive XC functional. In this way, we completely remove the correlation part of the interaction energy. We then add back the correlation energy by adding to the ``correlationless'' binding energy the non-additive correlation obtained with the GCP$^u_{1}$. The resulting binding energy expression defines the FDE-vdW method and is given by
\begin{equation}
\label{GCPbind}
E_{\rm bind}^{\rm FDE-vdW} = E_{\rm bind}^{\rm FDE}-E_{\rm c}^{\rm nadd}[\rho_I,\rho_{II}](GGA) +E_{\rm c}^{\rm nadd}({\rm GCP}^u_{1}).
\end{equation}

It is important to remark that the subsystem DFT formalism greatly
simplifies the theory by providing an algorithm that naturally includes
the non-additive correlation energy in the DFT energy expression, without invoking approximations in the underlying theory. Another important observation is that the theory so far ignores the relaxation of the subsystem
densities due to the non-additive, long range interaction. Since
these interactions are typically small for non-covalently bound
subsystems, this is a reasonable
approximation. 
\subsection{Computational details}
FDE and FDE-vdW binding energies are evaluated according to
\eqn{FDEbind}, and \eqn{GCPbind}, respectively. FDE and FDE-vdW
calculations are carried out employing the revPBE exchange density
functional \cite{revPBE} and PBE correlation \cite{PBEc} unless
otherwise stated. \eqn{dysonu} is solved within the adiabatic
approximation using the semi-local XC kernel corresponding to the
density functional employed. The FDE non-additive kinetic energy functional PW91k
\cite{lemb1994} is used  throughout. The calculations are carried out with a modified version of ADF \cite{adf}, and employ a TZ2P Slater-type orbital basis set unless otherwise stated.

All excitations were included in the sum in \eqn{sublr6b} with exception
of the large monomers (i.e.\ complexes containing a benzene monomer),
for which  excitations were sorted in descending order of contribution
to the molecular polarizability, similarly to  Ref.\
\cite{neug2010,neug2010a}. Only the excitations accounting for 99.5\% of
the isotropic, static polarizability were used to compute \eqn{sublr6b},
thereby reducing  computational cost while maintaining
a satisfactory level of accuracy. 
%In addition, to ensure a high accuracy in the fit of \eqn{sublrADF}, an augmented fit set for carbon atoms was used throughout.
%
%
\section{Results and discussion}
\subsection{FDE vs.\ FDE-vdW binding energies}
We calculated the binding energies of a selected set of weakly bound
complexes from the S22 set \cite{s22} using GCP$^u_{1}$. 
The S22 set is widely used to
benchmark the performance of methods. It consists of a set of dispersion
bound systems, a set of hydrogen-bound systems, and a set which is both
dispersion and hydrogen bound. Because the algorithm, as
implemented, scales as $o_I^3v_I^3+o_{II}^3v_{II}^3$, we are limited to apply it to
relatively small test cases. We therefore only selected complexes from
the S22 set with less than 20 atoms as well as the two benzene dimer
structures, one with stacked monomers
[(C$_6$H$_6$)$\cdots$(C$_6$H$_6$)$_\parallel$] and one T-shape
[(C$_6$H$_6$)$\cdots$(C$_6$H$_6$)$_\perp$]. This choice narrowed the set
to 13 complexes out of 22. A summary of the resulting binding energies
is found in Table \ref{tres}.
\begin{table}
\begin{center}
\begin{tabular}{lrrrr}
Complex & FDE(1) & FDE-vdW(1) & FDE-vdW(2) & CCSD(T)/CBS$^a$ \\
\hline
% small mols here
(NH$_3$)$_2$                                 &  $-$2.41  & $-$3.34  & $-$3.34    & $-$3.13 \\
(H$_2$O)$_2$                                 &  $-$3.39  & $-$4.28  & $-$4.25    & $-$4.99 \\
(HCOOH)$_2$                                  & $-$12.32 & $-$16.70 & $-$16.54   & $-$18.75 \\
(HCONH$_2$)$_2$                              & $-$12.73 & $-$16.77 & $-$16.74   & $-$16.06 \\
(CH$_4$)$_2$                                 &  $+$0.01  & $-$0.58  & $-$0.58    & $-$0.53 \\
(C$_2$H$_4$)$_2$                             &  $+$0.07  & $-$1.28  & $-$1.27    & $-$1.47 \\
(C$_2$H$_4$)$\cdots$(C$_2$H$_2$)             &  $-$1.01  & $-$1.66  & $-$1.67    & $-$1.50 \\
% large mols here                           
(C$_6$H$_6$)$\cdots$CH$_4$                   & $+$0.24  & $-$1.74   & $-$1.73    & $-$1.45 \\
(C$_6$H$_6$)$\cdots$H$_2$O                   & $-$1.04  & $-$3.39   & $-$2.58    & $-$3.28 \\
(C$_6$H$_6$)$\cdots$NH$_3$                   & $-$0.35  & $-$2.92   & $-$2.07    & $-$2.31 \\
(C$_6$H$_6$)$\cdots$HCN                      & $-$1.67  & $-$4.55   & $-$3.83    & $-$4.54 \\
(C$_6$H$_6$)$\cdots$(C$_6$H$_6$)$_\parallel$ & $+$0.89  & $-$2.64   & $-$2.15    & $-$2.65 \\
(C$_6$H$_6$)$\cdots$(C$_6$H$_6$)$_\perp$     & $+$0.34  & $-$1.88   & $-$2.00    & $-$2.72 \\
\hline
{\bf MUE}                                   & {\bf 2.31} & {\bf 0.46} & {\bf 0.57} & \\
\end{tabular}
\caption{\label{tres}Binding energies of 13 complexes from the S22 set.  MUE stands for mean unsigned error. Values in kcal/mol. In the FDE-vdW(2) calculations, the $\chi_I^u$ response functions appearing in $E^{\rm nadd}_c$ are calculated with the asymptotically corrected SAOP functional \cite{grit2000}.}
$a$ from Ref.\ \citenum{mars2011}
\end{center}
\end{table}
The results presented in Table \ref{tres} indicate that the FDE-vdW
method is consistently closer to the benchmark than semilocal FDE. The MUE is
decreased from 2.31 kcal/mol when the non-additive correlation is
evaluated with PBEc, to 0.46 and 0.57 when the GCP$^u_1$
correlation is evaluated either with the revPBE or the SAOP model
potential, respectively. 

We now discuss the effect of the asymptotic behavior of the XC
potential in the TD-DFT calculations. In the ideal case of employing a
complete basis set, the GCP$^u_{1}$ non-additive correlation
energy overestimates the correlation energy
computed with \eqn{sublr3} for two reasons: (1) GGA functionals tend to
underestimate excitation energies, which yields an overestimation of the
non-additive correlation energy, since
\eqn{sublr8}
contains excitation energies in the denominator, 
and (2) from arguments
related to the approximate coupling-strength integration (see the
appendix and Figure \ref{appfig}). Point (1) is known and currently
dealt with in the literature for implementations of the GCP formula by
employing hybrid XC potentials  in combination with asymptotic
corrections \cite{pode2012,pern2009b,misq2005,misq2003,misq2002}. 
We also observe that in moving from
revPBE to the asymptotically corrected SAOP model potential \cite{grit2000,schi2000,grit1999}, 
the FDE-vdW binding energies generally decrease in magnitude. 

Since the calculations were performed with a medium-sized basis set (the TZ2P
set, see next section for a basis set analysis), which truncates the
sum-over-states formula in \eqn{sublr6b}, we are bound to profit from 
cancellation of error -- although GCP$^u_1$ should
overestimate the correlation energy interaction, the finite basis set
used will offset such overestimation and possibly result in an
underestimation. The choice of basis set is analyzed in the following section.
\subsection{Choice of basis set}
It is well-known \cite{eshu2012} that correlation energies calculated from
RPA or TD-DFT response functions are significantly dependent on the
choice of
basis set. In the case of the GCP method, there is consensus that
the larger the basis set used, the larger the calculated dispersion
interaction \cite{misq2002}. This is what we also find for the
GCP$_1^u$ method as exemplified by the calculations presented
in Table \ref{tbas}.
\begin{table}
\begin{center}
\begin{tabular}{rrrrrr}
System                                          & (NH$_3$)$_2$ & (H$_2$O)$_2$ & (HCOOH)$_2$ & (HCONH$_2$)$_2$ & (C$_2$H$_4$)$\cdots$(C$_2$H$_2$) \\
\hline
$E_{\rm c}^{\rm nadd}$(TZ2P)  & $-$1.57           &  $-$1.64            & $-$7.13          & $-$5.97                   & $-$0.99 \\
$E_{\rm c}^{\rm nadd}$(QZ3P) & $-$1.80          & $-$2.01             &  $-$8.65          & $-$7.12                  & $-$1.35 \\
\# functions (TZ2P)      & 132                  & 108                     & 228                  & 252                       & 192 \\
\# functions (QZ3P)     &176                   & 144                     & 304                 & 336                         & 256      \\
%\hline
%{\bf \% energy deviation}                & {\bf 13}                    & {\bf 18}                     & {\bf 18}                    & {\bf 16}                          & {\bf 27}
\end{tabular}
\end{center}
\caption{\label{tbas} GCP$_1^u$ non-additive correlation energy contribution to the FDE-vdW interaction energy. Calculations carried out with the revPBE functional. The QZ3P basis set is a non-relativistic valence quadruple zeta basis set + 3 polarization functions \cite{ET-QZ3P}. Energy values in kcal/mol.}
\end{table}
This behaviour can be rationalized by looking at \eqn{sublr8}. The
larger the basis set, the more excitations one needs to include in the
sum and thus a larger non-additive correlation is found, up to an
asymptote. Assuming that the QZ3P calculations are close to the basis
set limit (as we use Slater-Type Orbitals featuring correct exponential decay and electron--nucleus cusps, faster convergence of the calculations with respect to basis set size is expected than for Gaussian-Type functions),  the values in Table \ref{tbas} show that already the TZ2P
set provides most of the non-additive correlation energy. Since only the
long-range non-additive part of the correlation energy is calculated using GCP$_1^u$, we
expect much less dependence on basis set size than for full RPA
correlation energies. This is also observed in so-called range-separated
RPA schemes \cite{toul2009,toul2004}.

Upon comparison of the binding energies in Table \ref{tres}, it is clear
that the employment of the smaller TZ2P basis set artificially improves
the FDE-vdW binding energies over the basis set limit due to the
aforementioned error cancellation. Similar effects are reported also for
the RPA method \cite{eshu2012}.

\subsection{Choice of functional}
Some approximate exchange functionals, such as the PW91 and PBE functionals, may artificially display minima for van der Waals complexes \cite{klim2010}. Others, such as the revPBE and the BP88, do not feature minima for van der Waals complexes and are often chosen to be the exchange counterpart of van der Waals density functionals \cite{vydr2011,dion2004,klim2010}.  
\begin{figure}[htp]
\begin{center}
\includegraphics[width=0.81\textwidth]{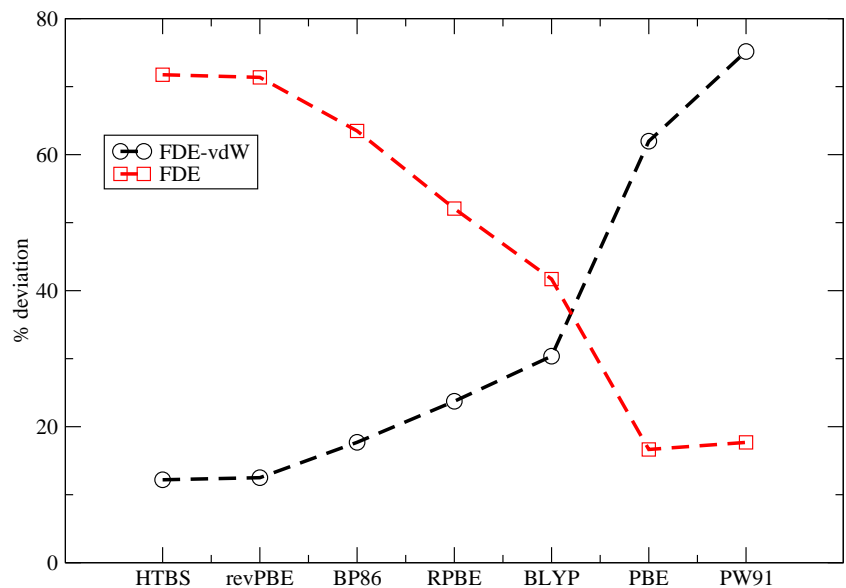}
\end{center}
\caption{\label{fig:funcs} Percent deviation of the FDE and FDE-vdW binding energies against the benchmark \cite{mars2011} when using the functional specified on the $x$-axis as the XC functional in the FDE calculations. In red (squares) the GGA functional was used to evaluate the full non-additive XC functional, while in black (circles) the GGA is employed for the exchange part and the GCP$_1^u$ is employed for the non-additive correlation part.}
\end{figure}

In this work, we carry out a preliminary analysis of the GGA XC
functional used in the FDE calculations, which then enters the
FDE/FDE-vdW binding energy expression in \eqs{FDEbind}{GCPbind}. The
binding energy of the van der Waals complexes featured in Table
\ref{tres} was recalculated employing 7 XC density functional
approximants. The results are summarized in Figure \ref{fig:funcs}. In
the figure, it is evident that PW91, PBE, BLYP, and RPBE introduce
spurious attractive interaction between the subsystems when they are
employed as non-additive exchange functionals. For this reason, they are
not suitable for pairing with a van der Waals correction, as previously
noted for KS-DFT calculation \cite{vydr2011,dion2004,pern2009b} and
here is also noted in FDE-vdW calculations. 

To conclude the XC functional analysis, the appropriate functionals for
pairing with the GCP$_1^u$ among the tested ones are: HTBS,
revPBE, and BP86. In a future study, we will carry out a complete
benchmark and a larger set of GGA functionals will be considered.

\section{Conclusions}
We have developed a new van der Waals inclusive DFT theory in the framework of subsystem DFT. Subsystem DFT is an ideal tool for identifying long and short range contributions to the binding energy, as they are associated respectively with the value of the additive and non-additive energy functionals. This is an important feature of the present theory which completely avoids the common parametrization of density functionals to classify short and long range interactions. Such parametrizations are often used to expedite the computational evaluation of van der Waals interactions \cite{toul2009}.

As a result of this work, the needed equations to evaluate van der Waals interactions arising between overlapping electron densities are now available. This has been an outstanding issue in this field, as a DFT theory for van der Waals interactions between two distinct electron densities has been available only in the non-overlapping case \cite{dobs2012}. 

A recent energy decomposition analysis of subsystem DFT calculations \cite{beyh2013} has shown that the inclusion of empirical van der Waals terms in the interaction energy has the effect of improving the binding energies calculated with GGA functionals. In line with those findings, we show that the FDE binding energies are dramatically improved if the non-additive correlation energy functional is made non-local in character and is evaluated with the Fluctuation--Dissipation Theorem. 

Preliminary results are promising and lead us to expect that FDE-vdW can deliver good binding energies across three binding regimes: dispersion, dipole, and mixed dispersion-dipole. 

A very appealing property of FDE-vdW is its potential for systematic improvement. For example, by including a numerical integration over the coupling strength parameter, $\alpha$, which would fold in non-local interaction effects beyond RPA. Another interesting improvement is to use the second term in the perturbative expansion in \eqn{sublr4}. In this way, three-body terms would be included in the overall non-additive correlation.

There are two clear weak points in the current implementation of the theory. First, the computational scaling for evaluating the non-additive correlation is prohibitive and it is typical of a correlated wavefunction method. In addition, the exact diagonalization of the Casida equations needed for the evaluation of the subsystem response function is only possible for small systems. Larger systems, such as uracil, indol and pyrazine, incur into linear dependencies of the $\{\phi_i\phi_a\}$ occupied-virtual product space that prevents a numerically stable diagonalization. The second weak point is that, in subsystem DFT, kinetic energy functionals are needed (at the non-additive level) for allowing a proper description of overlapping electron densities. The available kinetic energy functionals are mostly formulated at the GGA level \cite{lari2014,lemb1994,jaco2007,jaco2013} which are known to describe well only certain types of weak interactions and not others \cite{kevo2006,gotz2009} (although GGA XC functionals had been used in the referred studies). 

Thus, the present theory will need to be accompanied in the future with more accurate kinetic energy density functionals, and more computationally amenable procedures for evaluating the response functions and the associated non-additive correlation energy. Efforts to address both points are ongoing in our group and are based on the formulation of non-local kinetic energy functionals to address the former issue and on a density fitting approach to address the latter which will be inspired by the techniques developed by Szalewicz and coworkers \cite{pode2012}. 

\section*{Acknowledgements}
We are grateful to Alisa Krishtal for the feedback provided on an early version of the manuscript.
Acknowledgment is made to the Donors of the American Chemical Society Petroleum Research Fund and the office of the Dean of FASN of Rutgers-Newark for partial support of this research.

\appendix
\section{Justification of the GCP$^u_1$ approximation}
In this section, we will show that the integrand of \eqn{sublr3b} after all integrals are carried out with the exception of the one over the coupling strength, $\alpha$, has positive and approximately monotonically decreasing derivative. Thus, if the response functions are chosen at a fixed value of $\alpha$, a good choice is $\alpha=1$.
\subsection{Derivative of the correlation energy with respect to the coupling strength}
Let us define $\epsilon_{\rm c}^{\rm nadd}(\alpha)$ as the value of the integrand in the definition of $E_c^{\rm nadd}$ in \eqn{sublr3b}  when the imaginary frequency and the spatial integrations are carried out. Consider the first derivative of the integrand of \eqn{sublr3b}, contracting the frequency and the spatial dependence, 
\begin{equation}
\label{a1}
\frac{\d \left[\chi_I^{u,\alpha} K_{IJ}^{\alpha} \chi_J^{u,\alpha}\right]}{\d\alpha}=\frac{\d \chi_I^{u,\alpha}}{\d\alpha}K_{IJ}^{\alpha} \chi_J^{u,\alpha}+\chi_I^{u,\alpha} K_{IJ}^{\alpha} \frac{\d \chi_J^{u,\alpha}}{\d\alpha} + \chi_I^{u,\alpha} \frac{\d  K_{IJ}^{\alpha}}{\d\alpha} \chi_J^{u,\alpha}.
\end{equation}
Realizing that 
\begin{equation}
\label{a2}
\frac{\d \chi_I^{u,\alpha}}{\d\alpha}= \chi_I^{u,\alpha} \frac{\d  K_{IJ}^{\alpha}}{\d\alpha} \chi_I^{u,\alpha} \simeq \chi_I^{0} \frac{\d  K_{IJ}^{\alpha}}{\d\alpha} \chi_I^{u,\alpha} 
\end{equation}
and evaluating \eqn{a1} in the RPA approximation, i.e.\
\begin{equation}
\label{a3}
K_{IJ}^{\alpha}= K_{II}^{\alpha} =K_{JJ}^{\alpha}=\frac{\alpha}{|\br-\brp|},
\end{equation}
leads to 
\begin{equation}
\label{a4}
\frac{\d \left[\chi_I^{u,\alpha} K_{IJ}^{\alpha} \chi_J^{u,\alpha}\right]}{\d\alpha}\simeq\left(\chi_I^{u,\alpha} - \chi_I^{0}  \right)\frac{\alpha}{|\br-\brp|} \chi_J^{u,\alpha}+\chi_I^{u,\alpha} \frac{\alpha}{|\br-\brp|} \left(\chi_J^{u,\alpha} - \chi_J^{0}  \right) + \chi_I^{u,\alpha} \frac{1}{|\br-\brp|} \chi_J^{u,\alpha},
\end{equation}
which could have been achieved by simple evaluation of the derivative through finite differences of the response functions. 
Realizing that the terms involving the response function difference are one-order-of-magnitude smaller than the other term, the above equation leads to the following approximate derivatives
\begin{align}
\frac{\partial \epsilon_{\rm c}^{\rm nadd}(\alpha)}{\d\alpha}\simeq {\rm Tr}\left[\chi_I^{u,\alpha} \frac{\d  K_{IJ}^{\alpha}}{\d\alpha} \chi_J^{u,\alpha}\right].
\end{align}
In the above equation, by the short-hand notation ``Tr'' we mean integration over all variables with exception of $\alpha$.
\subsection{Linear extrapolation of  $\epsilon_{\rm c}^{\rm nadd}(\alpha)$}
\begin{figure}
\begin{center}
\includegraphics[width=8cm]{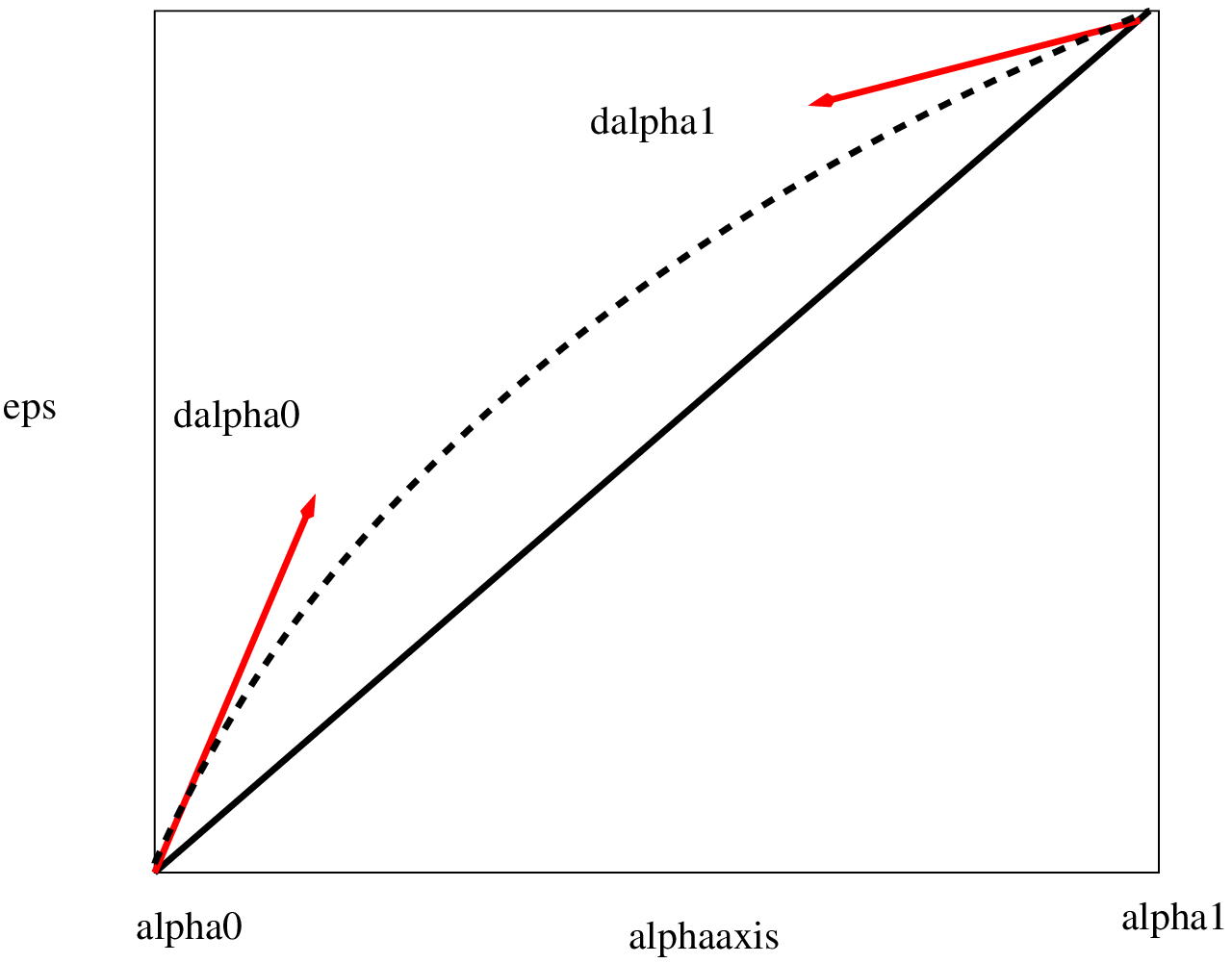}
\end{center}
\caption{\label{appfig}Predicted trend of the integrand in \eqn{sublr3b} as a function of the coupling stregth parameter.}
\end{figure}

Recalling \eqn{a3}, and also that the evaluation of the integral in \eqn{sublr3b} with the KS response functions ($\chi^0_I$) yields much larger correlation energies than with correlated response functions \cite{hess2003}, we find the following approximate inequality
\begin{equation}
\label{a5}
\left.\frac{\d\epsilon_{\rm c}^{\rm nadd}}{\d\alpha}\right|_{0}\geq \left.\frac{\d\epsilon_{\rm c}^{\rm nadd}}{\d\alpha}\right|_{\alpha}.
\end{equation}
Because the integrand of \eqn{sublr3b} at zero coupling strength is exactly zero, the trend of $\epsilon_{\rm c}^{\rm nadd}(\alpha)$ can be inferred. Namely, $\epsilon_{\rm c}^{\rm nadd}(\alpha)$ always lies above the linear extrapolation line, see Figure \ref{appfig}. For comparison, see Figure 3 in Ref.\ \citenum{furc2008} keeping in mind that the plot there is inverted on the $y$-axis when comparing it to our Figure \ref{appfig}.
\psfrag{alpha0}{$\alpha=0$}
\psfrag{alpha1}{$\alpha=1$}
\psfrag{alphaaxis}{$\alpha$}
\psfrag{eps}{$\epsilon_{\rm c}^{\rm nadd}$}
\psfrag{dalpha0}{$\left.\frac{\d\epsilon_{\rm c}^{\rm nadd}}{\d\alpha}\right|_{0}$}
\psfrag{dalpha1}{$\left.\frac{\d\epsilon_{\rm c}^{\rm nadd}}{\d\alpha}\right|_{1}$}
The above analysis, shows that a good approximation to the true $\epsilon_c^{\rm nadd}(\alpha)$ is simply the linear extrapolation, which is realized by \eqn{subecorr}, using the RPA approximation. 

From the graph, it is also inferred that the linear extrapolation of the integrand using as slope the derivative at a fixed $\alpha$ will yield overestimated correlation energies, and that the largest overestimation is achieved when $\alpha=0$, and a much reduced overestimation is found if $\alpha=1$.

\end{document}